\documentclass[floatfix, superscriptaddress, reprint, prl,aps,10pt]{revtex4-1}
\usepackage{CJK}
\usepackage[utf8]{inputenc}
\usepackage{amsmath}
\usepackage{amsfonts}
\usepackage{amssymb,amsthm}
\usepackage[subrefformat=parens,labelformat=parens,caption=false]{subfig}
\usepackage{multirow}
\usepackage[]{hyperref}
\usepackage{bbold}
\usepackage{cancel}
\newcommand{\eq}[1]{\begin{align}#1\end{align}}
\hypersetup{
  colorlinks = true,
  urlcolor = magenta,
  linkcolor = red,
  citecolor = blue
}

\usepackage[all]{hypcap}
\usepackage{url}
\usepackage{xcolor}
\newcommand{\ccol}[2]{\color{#1}{{#2}}\color{black}}
\usepackage[normalem]{ulem}
 \usepackage{subfig}

\usepackage{graphicx}

\begin{document}

\title{Single-photon bound states in atomic ensembles}

\author{Yidan Wang}
\affiliation{Joint Quantum Institute, NIST/University of Maryland, College Park, MD 20742, USA}	
\author{Michael J. Gullans}
\affiliation{Department of Physics, Princeton University, Princeton, NJ 08540, USA}
\author{ Antoine Browaeys}
\affiliation{Laboratoire Charles Fabry, Institut d'Optique Graduate School, CNRS, Universit\'{e} Paris-Saclay, 91127 Palaiseau cedex, France}
\author{ J. V. Porto}
\affiliation{Joint Quantum Institute, NIST/University of Maryland, College Park, MD 20742, USA}
\author{Darrick E. Chang}
\affiliation{ICFO-Institut de Ciencies Fotoniques, Mediterranean Technology Park, 08860 Castelldefels (Barcelona), Spain}
\affiliation{ICREA-Instituci\'o Catalana de Recerca i Estudis Avan\c{c}ats, 08010 Barcelona, Spain}
\author{Alexey V. Gorshkov}
\affiliation{Joint Quantum Institute, NIST/University of Maryland, College Park, MD 20742, USA}
\affiliation{Joint Center for Quantum Information and Computer Science, NIST/University 
of Maryland, College Park, MD 20742, USA}

\begin{abstract}
We illustrate the existence of single-excitation bound states for propagating photons interacting with $N$ two-level atoms. These bound states can be calculated from an effective spin model, and their existence relies on dissipation in the system. The appearance of these bound states is in a one-to-one correspondence with  zeros in the single-photon transmission  and with divergent  bunching in the second-order photon-photon correlation function. We also formulate a dissipative version of  Levinson's theorem for this system by looking at the relation between the number of bound states and the winding number of the transmission phases. This theorem allows a direct experimental measurement of the number of bound states using the measured  transmission phases. 
\end{abstract}

\maketitle

Systems of strongly interacting photons are important for scaling up quantum computers and networks  \cite{Lodahl2017}. The generation and manipulation of non-classical light enable  single-photon switches and transistors \cite{Chang2007, Tiarks2014,  Bajcsy2009}, quantum circulators \cite{Scheucher2016}, isolators \cite{Sayrin2015}, and long-distance quantum state transfer \cite{Vermersch2017,Yang2017}.  Understanding few-body physics in systems of strongly interacting photons helps reveal the emergent many-body physics in these systems \cite{Moos2015}, including quantum phase transitions of light \cite{Greentree2006,Fink2017}.

To characterize the quantum states of light produced in these systems, experiments often probe single-photon transmission and multi-photon statistics.
In this Letter, we construct an effective spin model to solve for the single-photon transmission through ensembles of atoms exhibiting cooperative light scattering effects.
This spin model is characterized by the presence of single-photon bound states, whose wavefunction is a hybridized single-excitation between light and matter that is localized in space. 
The bound states correspond directly to zeros in the transmission coefficient and are associated with an analogue of Levinson's theorem \cite{Wellner1964, Wright1965,Dong2000}  for interacting atom-photon  systems.  In the two-photon transmission, we show that these zeros in the transmission lead to divergent  bunching of the light due to the presence of effective photon-photon interactions  in these systems.

Our results and their generalizations in a following paper \cite{upcoming} can be applied to many systems in which strong interaction between propagating photons are realized. These include photonic-crystal waveguides coupled to atoms \cite{Hung2013,Goban2014, Goban2015, Hood2016, Angelatos16} and solid-state emitters \cite{Hausmann2012,Sipahigil2016Nov,Lodahl2015},  free-space photons propagating through dense Rydberg atomic clouds \cite{Peyronel2012, Firstenberg2013}, optical nanofibers coupled to atoms \cite{ Goban2012,  Sayrin2015, Scheucher2016, Sorensen2016, Corzo2016,Solano2017,Solano2017a}, transmission lines coupled to superconducting qubits \cite{VanLoo2013, Devoret2013}, and metallic nanowires coupled to NV centers \cite{Huck2011} and quantum dots \cite{Akimov2007}.  These systems typically consist of freely propagating photons coupled to a number of emitters that provide the nonlinearity either individually or due to inter-emitter interactions. Recent years have seen significant developments in theoretical methods for such systems  \cite{Shi2009, Fan2010, Pletyukhov2012, Zheng2012, Lalumiere2013, Baragiola2014, Shi2015m, Caneva2015, Xu2015a,  Asenjo-Garcia2017, Roy2017, Das2018, Das2018a}. In several of these approaches, the emitters' dynamics are characterized by a spin model $M^{{\rm tot}}$, which consists of the energy of the atoms and the photon-mediated dipolar interactions $K^{{\rm tot}}$ between them. $K^{{\rm tot}}$ contains both a coherent and a dissipative part.  For two atoms in vacuum separated by distance $r$, $K^{{\rm tot}}$ increases as $1/r^3$ at small $r$ \cite{Chomaz2012}.   Photonic observables are then related to $M^{{\rm tot}}$ in an indirect way via an input-output relation involving emitter degrees of freedom \cite{ Lalumiere2013, Caneva2015, Xu2015a, Shi2015m}. In contrast, we construct here a spin model $M$ that directly encodes the transmission properties.

 We provide a concrete analysis of two-level atoms interacting with linear-dispersion photons, but our approach readily generalizes to other level structures and dispersion relations. \cite{upcoming}.
\begin{figure}[]
\includegraphics[width=0.7\linewidth]{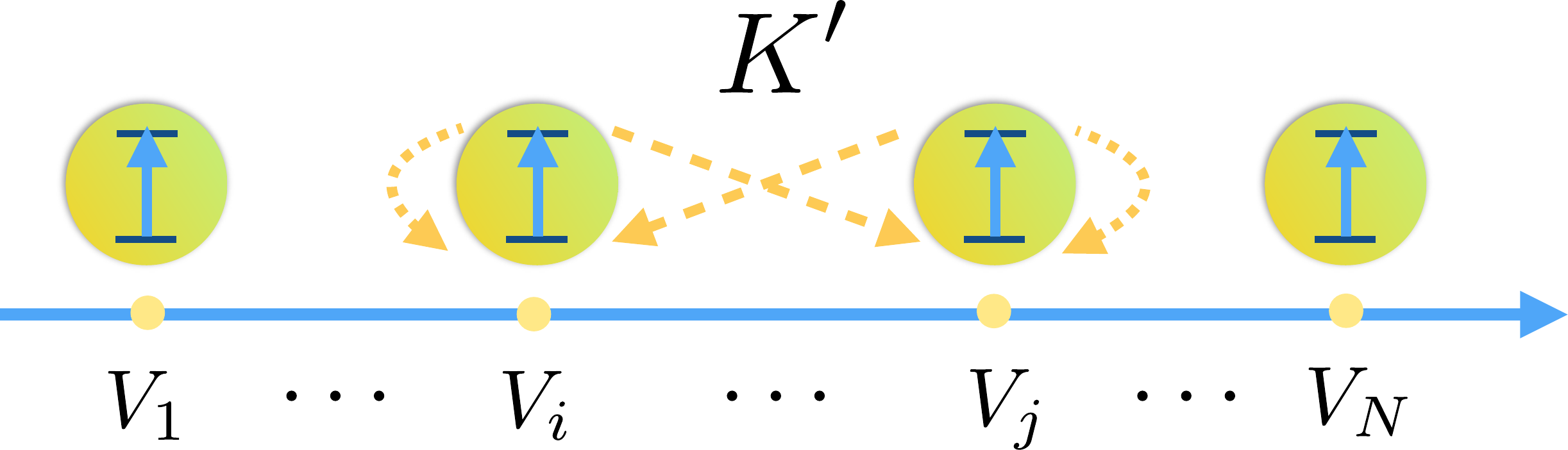}
\caption{(Color online) An illustration of $N$ atoms coupled to a 1D photon channel, with $V_i$ corresponding to the amplitude of the coupling of the photons to atom $i$. In addition to this specific photonic channel of interest, the atoms can also interact via additional photonic channels, which results in an effective dipole-dipole interaction $K'$. }
\label{fig_keynote}
\end{figure}
 Figure \ref{fig_keynote} illustrates the system we consider. Photons propagate to the right through a 1D channel and interact with $N$ two-level atoms with energy $\omega_{eg}$.
 In addition to the 1D channel of interest, the atoms also interact with other photonic modes called the reservoir modes. 
If we are not interested in scattering into these other modes, their effect can be captured by a dipole-dipole interaction $K'_{ij}$, which physically describes photon emission and re-absorption via the other modes.   Note that if $K$ describes the additional dipole-dipole interactions mediated by the 1D channel, $K^{{\rm tot}}=K'+K$.  In many nanofiber experiments, photons are sent into the fiber in a particular transverse mode and measured at the output in the same transverse mode. Such systems are well described by our model, where the 1D channel consists of a continuous spectrum of right-propagating photons in the transverse mode of interest, while the reservoir modes include left-propagating photons and photons in other transverse modes. 
 
The 1D system is described by the master equation
\eq{
 \dot{\rho}=-i(H\rho-\rho H^{\dagger})+i\sum_{i,j=1}^{N} (K'_{ij}-K^{'\dagger}_{ij})b_i\rho b^{\dagger}_j,\label{eq_MasterEquation}
}
where the effective Hamiltonian $H$ is non-Hermitian and the recycling term (the last term) comes from the reservoir-induced interaction $K'$. We treat the two-level atoms as harmonic oscillators with creation operators $b_i^\dagger$ and introduce a hard-core interaction $U$ in $H$ to prevent atoms from being excited more than once. Hence $H$ is given by 
  \begin{gather}
  H=H_0+V+U,\\
H_0=-\int_{-\infty}^{+\infty} dz\ i C^\dagger(z)\partial_{z}C(z)+\sum_{i=1}^{N} \omega_{eg}b^{\dagger}_ib_i, \\
V= \left[\sum_{i=1}^NV_iC(z_i)  b^{\dagger}_{i}+\text{h.c}\right]+\sum_{i,j=1}^NK'_{ij}b_{i}^{\dagger}b_{j}, \\
U=\sum_{i=1}^{N}ub_i^{\dagger}b_i^{\dagger}b_ib_i, \quad \quad u=+\infty,
\end{gather} 
 where $H_{0}$ is the Hamiltonian of the right-propagating free photons with linear dispersion (speed of light $c=1$) and $N$ noninteracting two-level atoms. $C^\dagger(z)$ creates a photon in the transverse mode of interest at position $z$ in the 1D channel, and $V$ is the quadratic interaction. The first term in $V$ describes the atom-photon interaction whereby atom $i$ is excited through the absorption of a photon at location $z_i$. The second term describes the dissipative dipole-dipole interaction $K'$ between atoms induced by the Markovian reservoir. Its Hermitian component describes coherent dipolar interactions while the anti-Hermitian component describes collective spontaneous emission.
In our definition, $K'$ is dissipative if $-i(K'-K'^\dagger)$ has only non-positive eigenvalues. 
Since we are interested in  scattering amplitudes  where the number of output photons in the 1D channel is the same as the number of input photons,  it suffices to consider the effective Hamiltonian $H$ in place of the full master equation \cite{Peyronel2012b,Gorshkov2013}.

\emph{Dissipative bound states.---}Consider the eigenstates of the effective Hamiltonian $H$ in the single-excitation Hilbert space, whose projected Hamiltonian is denoted as $H_1$. Note that $H_1$ also describes the classical problem where coherent states of light interact with classical dipoles. Therefore, all the single-photon results can be applied to this classical problem as well.

Every incoming free photon $|k\rangle$ uniquely labels a scattering state with energy $k$; therefore, the scattering-state energy spectrum is $(-\infty,+\infty)$. There may also exist bound states in the continuum \cite{vonNeumann1929, Friedrich1985, Hsu2016}, but their existence requires fine tuning of parameters. When $K'$ describes coherent interaction, $H_1$ is Hermitian and its spectrum is real; there are no generic bound states in the system. When  $K'$ includes dissipative interactions, $H_1$ becomes non-Hermitian. In principle, we can have bound states with complex eigenenergies, isolated from the real line occupied by the scattering states.

The non-Hermitian nature of $H_1$ implies  that its left and right eigenvectors can be different and the set of all left (or right) eigenvectors is not guaranteed to form a complete basis. Generically the basis of eigenvectors is complete, in which case the \ccol{black}{Hamiltonian } is diagonalizable and takes the form
\begin{gather}
H_1=\int_{-\infty}^{+\infty} dk \ k|\psi_k\rangle \langle \pi_k|+\sum_{\alpha=1}^{N_B} E_{\alpha}|\psi_{\alpha}\rangle \langle \pi_{\alpha}| \label{eq_H1dec},
\end{gather}
where $|\psi_k\rangle$ and $|\pi_k\rangle$  are, respectively, the right and left scattering eigenvectors with eigenenergy $E_k=k$ and $|\psi_{\alpha}\rangle$ and $|\pi_{\alpha}\rangle$ are, respectively, the right and left bound states with complex eigenenergy $E_{\alpha}$. $N_B$ is the number of bound states. The orthogonality relations among the eigenstates are 
$
 \langle \pi_k|\psi_{k'}\rangle=\delta(k-k'),   \langle \pi_{\alpha}|\psi_{\beta}\rangle=\delta_{\alpha\beta},
  \langle \pi_k|\psi_{\alpha}\rangle=  \langle \pi_{\alpha}|\psi_k\rangle=0.
$ 
From a mathematical perspective, the bound states are necessary for the completeness of the basis:  $
\mathbb{1}=\int_{-\infty}^{+\infty} dk  |\psi_k\rangle \langle \pi_k|+\sum_{\alpha=1}^{N_B} |\psi_{\alpha}\rangle \langle \pi_{\alpha}|\label{eq_compl}$.  It is one of our main goals in this work to understand the physical significance of these dissipative bound states.
The first main result of this Letter is that the bound state eigenenergies $E_{\alpha}$ can be calculated from an effective spin model $M$. For a single excitation, $M$ is an $N\times N$ matrix:
\begin{gather}
M_{ij}=\omega_{eg}\delta_{ij}+K'_{ij}+K^{\dagger}_{ij},\label{eq_M}
\end{gather}
where $K$ is the dissipative dipole-dipole interaction induced by the 1D photon modes of interest:
\begin{gather}
K_{ij}=-iV_iV^*_j \exp(i\omega_{eg}(z_i-z_j))\Theta(z_i-z_j),\label{eq_K1}
\end{gather}
and $K^\dagger$ is its Hermitian conjugate. In comparison, $M^{{\rm tot}}_{ij}=\omega_{eg}\delta_{ij}+K'_{ij}+K_{ij}$  is the  single-excitation effective Hamiltonian matrix for the atoms after all the photon channels are traced out of the system.  One can intuitively understand $K$ as describing a process dissipating energy from the atoms to the 1D channel, while $K^\dagger$ describes an energy-absorption process. In this context, the appearance of $K^\dagger$ in $M$ can be directly related to an energy conservation condition in the bound-state eigenvalue equation, as explained in the Supplemental Material. 

\begin{figure}
\subfloat[Eigenvalues of $M$]{
    \includegraphics[width=0.5\linewidth]{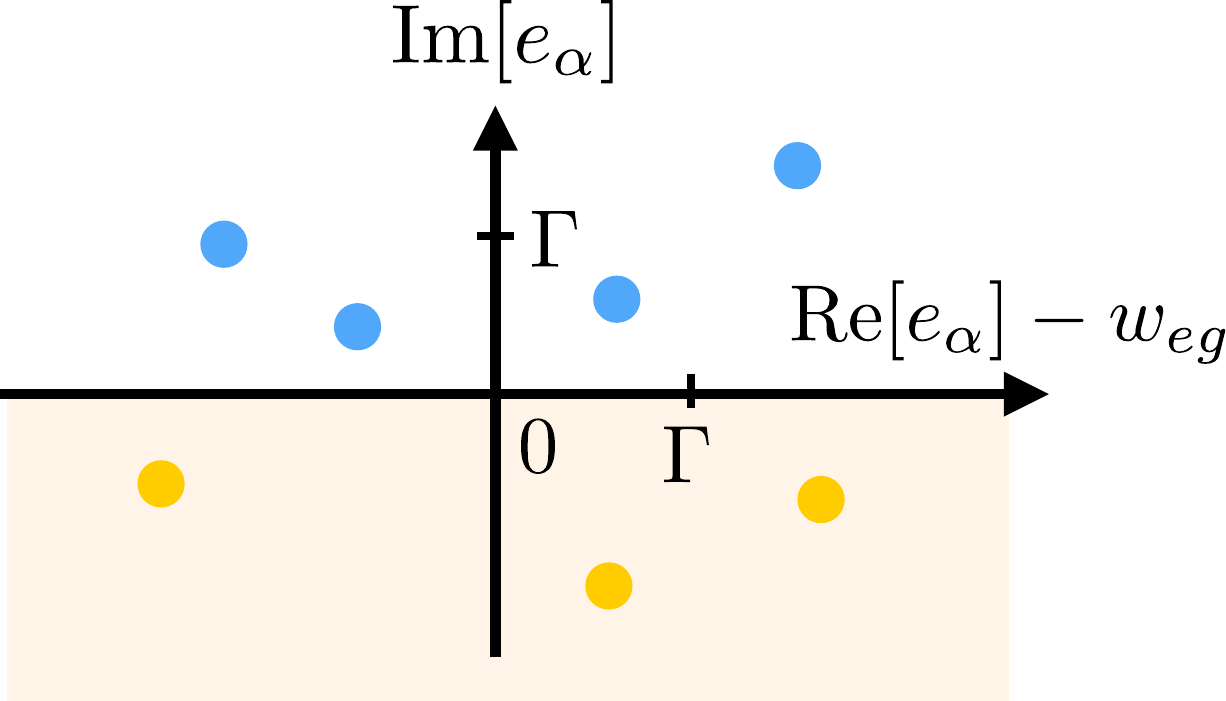} \label{fig:EigM}
}
 \subfloat[Eigenvalues of $M^{{\rm tot}}$]{
    \includegraphics[width=0.5\linewidth]{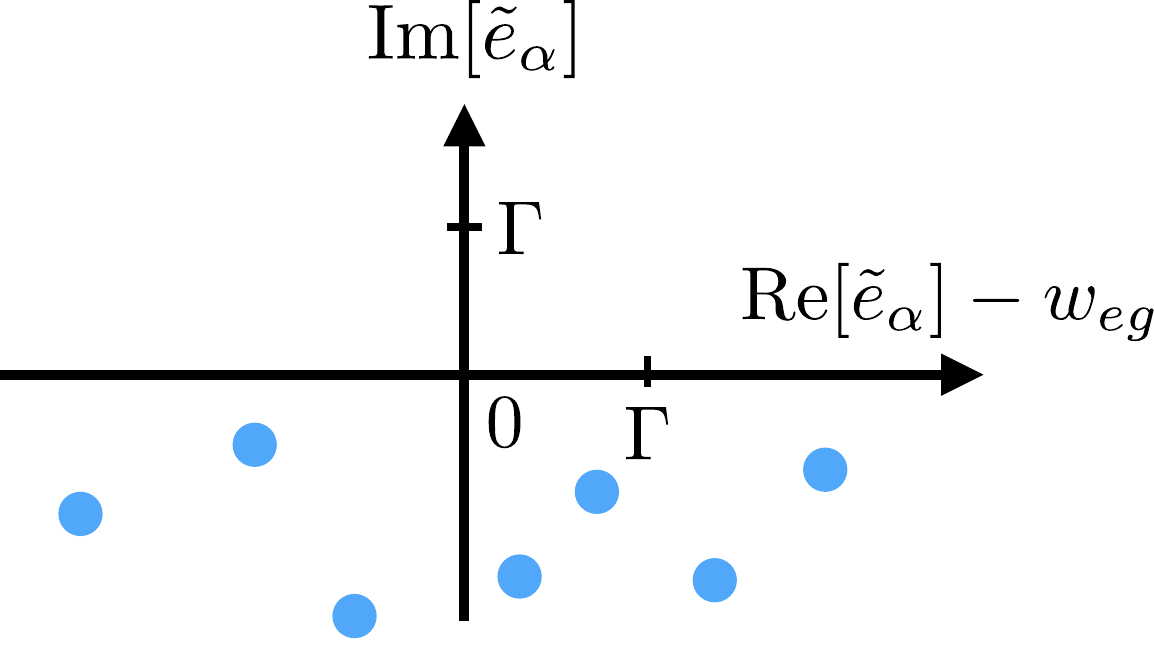} \label{fig:EigMtot}
}
\caption{(Color online) Schematic of the eigenvalues of (a) $M$ and (b) $M^{{\rm tot}}$ for $N=7$. $\Gamma \sim \frac{V^2_i}{2}$ is the scale of the single-atom decay rate to the 1D channel.  The eigenvalues of $M$ in the lower half plane (shaded region) give valid bound states (yellow). There is no eigenvalue of $M^{{\rm tot}}$ above the real line. }\label{fig:eigenvalues}
\end{figure}

 Because $M^{{\rm tot}}$ is dissipative, all its eigenvalues are either on or below the real line. Unlike $M^{{\rm tot}}$, the eigenvalues of $M$ can be anywhere with respect to the real axis; the ones below the real axis correspond to the bound state energies. Fig.\ \ref{fig:eigenvalues} shows schematically the eigenvalues of $M$ and $M^{{\rm tot}}$ for $N=7$. In this case, there are $N_B=3$ bound states illustrated as yellow dots.  These bound states have a hybrid light-matter wavefunction whereby the photonic field of the 1D mode is localized in space (i.e., bound) near the atoms. The bound states constantly leak energy into the reservoir modes without changing shape.

Given a bound-state energy $E_\alpha$, if its algebraic and geometric multiplicity are different as an eigenvalue of $M$, both $M$ and $H_1$ are non-diagonalizable. In this case, Eq.\ \eqref{eq_H1dec} and the completeness relation of the eigenstates of $H_1$ do not hold. However, one can define generalized bound states using so-called ``generalized eigenvectors'' of $M$ corresponding to $E_\alpha$. With these generalized bound states, the completeness relation is restored \cite{upcoming}.

\emph{Transmission coefficient $t_k$.---}The right scattering eigenstate with energy $k$ can be written as
\begin{gather}
|\psi_k\rangle=\int_{-\infty}^{+\infty}dz \phi _k(z)C^{\dagger}(z)|0,g\rangle +\sum_{j=1}^Ne_{j,k}b_j^\dagger|0,g\rangle,\label{eq_psik_def}
\end{gather}
where $\phi_k(z)$ is the photon wavefunction and $e_{j,k}$ is the excitation amplitude for the $j$-th  atom. $|0,g\rangle$ is the ground state with zero excited atoms or photons. Outside the region of atoms, $\phi_k(z)$ is a plane wave satisfying $\lim_{z\rightarrow \pm\infty}\phi_k(z)=a^{\pm}_{k}\exp(ikz)$. The transmission coefficient $t_k(z)=a^+_k/a^-_k$ is an easily accessible experimental observable.  
The second main result of this Letter is:
\begin{gather}
t_k=\frac{\text{det}(k\mathbb{1}-M)}{\text{det}(k\mathbb{1}-M^{{\rm tot}})}=\prod_{\alpha=1}^{N}\frac{k-e_{\alpha}}{k-\tilde{e}_{\alpha}},\label{eq_tk}
\end{gather}
where $\mathbb{1}$ is an $N\times N$ identity matrix, and $e_\alpha$ and $\tilde{e}_\alpha$ are the eigenvalues of $M$ and $M^{{\rm tot}}$, respectively.
In the literature, the transmission through the 1D channel is usually derived from $M^{{\rm tot}}$ via an input-output relation \cite{ Lalumiere2013, Caneva2015, Xu2015a, Shi2015m}. Here, however, we give a concise formula where a different spin model $M$ directly encodes this useful information about photon transport.  Eq.\ \eqref{eq_tk} is proved in the Supplemental Material \cite{SM_bs}.
 
The second equality in Eq.\ \eqref{eq_tk} is applicable when both $M$ and $M^{{\rm tot}}$ are diagonalizable. It shows that if a real eigenvalue $e_{\alpha}$ of $M$ is not an eigenvalue of $M^{{\rm tot}}$, then the transmission coefficient $t_{k=e_{\alpha}}=0$. The corresponding eigenstate is not a bound state\textemdash it is a scattering state with zero transmission.  When $e_{\alpha}$ crosses the real line in the complex plane during the continuous tuning of parameters, a bound state appears or disappears.
If a real eigenvalue $e_{\alpha}$ of $M$ is also an eigenvalue of $M^{{\rm tot}}$, it corresponds to a bound state in the continuum \cite{upcoming}.  In this case, $t_{k=e_{\alpha}}$ is not necessarily $0$. 

Consider the well-studied example of a single atom ($N=1$) at $z_1=0$ with $V_1=V$ and decay rates $K'=-i\Gamma'$ and $K=-iV_1^2/2\equiv -i\Gamma$  \cite{Shen2007,Rephaeli2013m}.   Here, $M=\omega_{eg}-i\Gamma'+i\Gamma$, $M^{{\rm tot}}=\omega_{eg}-i\Gamma'-i\Gamma$, resulting in $t_k=\frac{k-(\omega_{eg}-i\Gamma'+i\Gamma)}{k-(\omega_{eg}-i\Gamma'-i\Gamma)}$. Physically, $\Gamma$ and $\Gamma'$ correspond to the atomic decay rate into the 1D channel and reservoir modes, respectively. When $\Gamma'>\Gamma$, a bound state exists with eigenenergy $E_B=\omega_{eg}-i\Gamma'+i\Gamma$. Its wavefunction is given in the Supplemental Material \cite{SM_bs}.  In the special case with $\Gamma=0$, the atom is decoupled from the 1D photon channel, and the bound state is simply the atom in its excited state with eigenenergy $E_B=w_{eg}-i\Gamma'$ \cite{SM_bs}.   When $\Gamma'<\Gamma$, no bound state exists.  At $\Gamma'=\Gamma$, $t_{k=\omega_{eg}}$ is equal to $0$.  Thus, we see that the decay to the reservoir is the key to the existence of the bound states.

\begin{figure}
\subfloat[$N=1$]{
\includegraphics[width=0.44\linewidth]{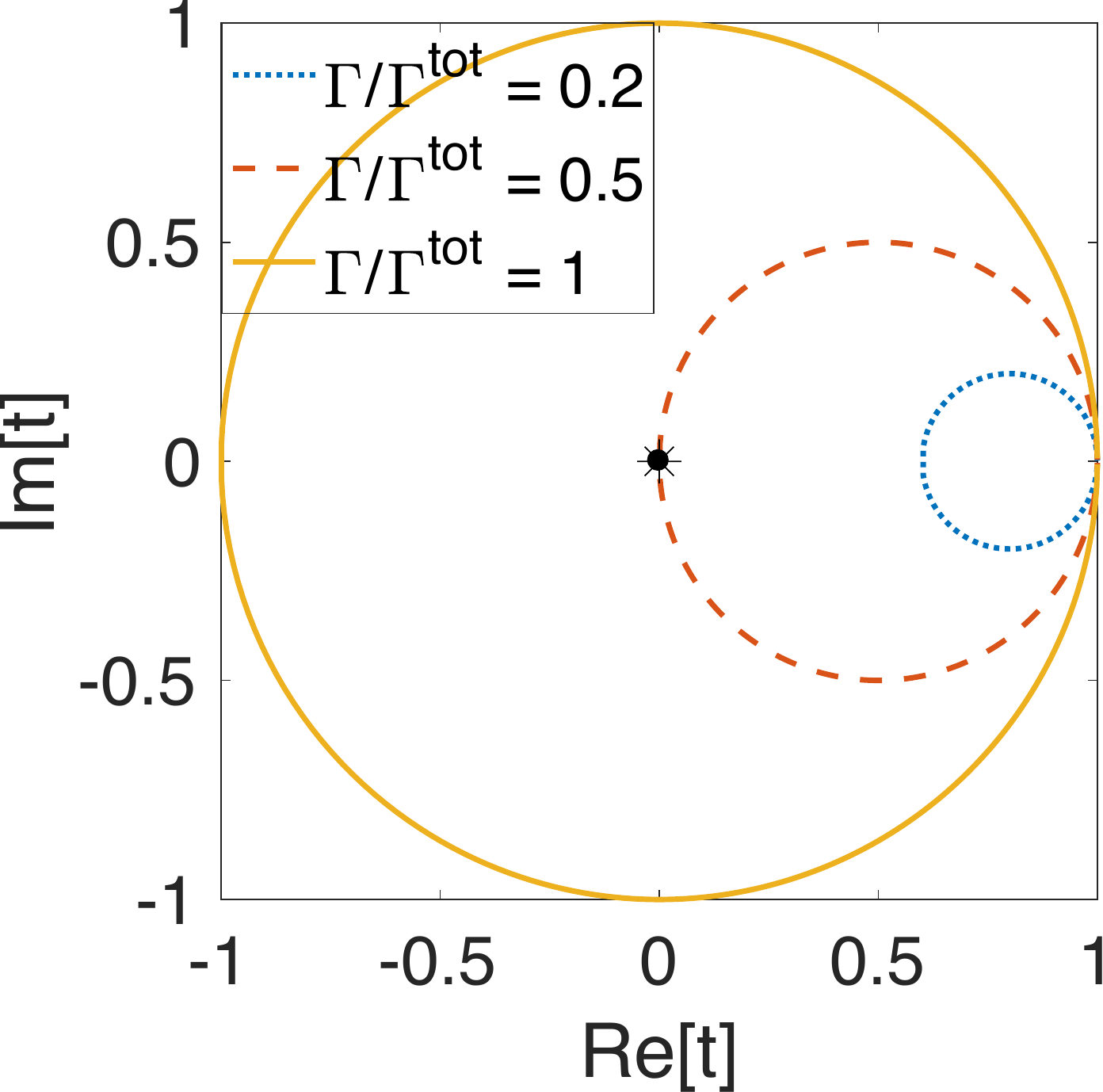} \label{fig_1a_tk}
}
\subfloat[$N=2$]{
\includegraphics[width=0.44\linewidth]{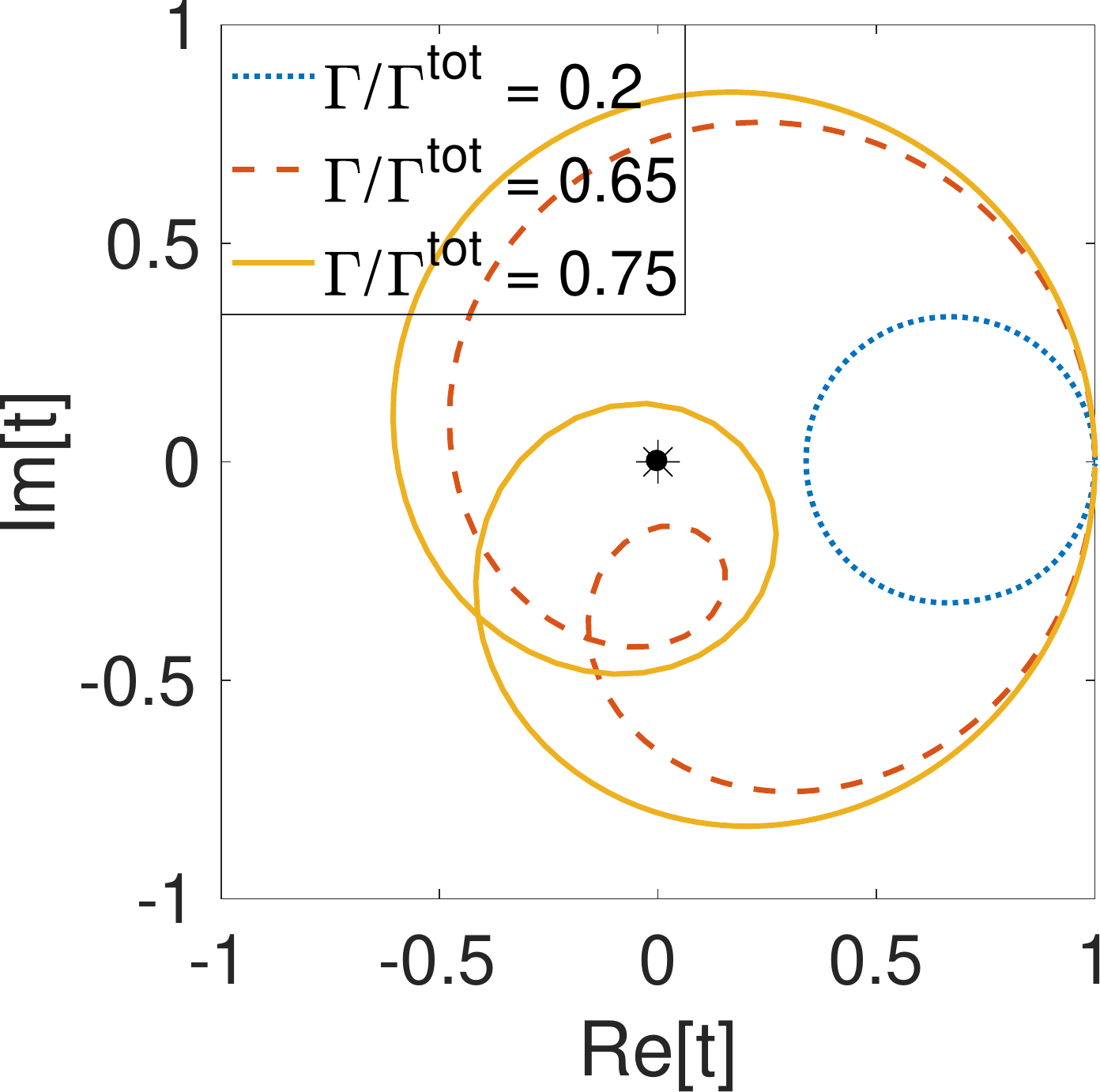} \label{fig_2a_tk}
}
\caption{(Color online) Im$[t_k]$ vs. Re$[t_k]$ as $k$ varies from $-\infty$ to $+\infty$ for (a) $N=1$ and (b) $N=2$. The trajectories start from $(1,0)$, go in the counter-clockwise direction, and end at $(1,0)$. The origins are marked as stars in the center of the figures. The number of times the trajectories enclose $(0,0)$ is equal to $N-N_B$, where $N_B$ is the number of dissipative bound states.}
\end{figure}

 \emph{Levinson's theorem.---}Since the bound states are orthogonal to the scattering states, they cannot be excited by sending photons into the 1D channel, which raises the the question: How can one probe these bound states in a scattering experiment? We recall Levinson's theorem for the Schr\"{o}dinger equation,  which relates the number of bound states and the difference of scattering phase shifts at zero and infinite energy \cite{Wellner1964, Wright1965,Dong2000}.  If a similar relation exists in our interacting atom-photon  system, then we can measure $N_B$ directly in experiments through a phase measurement of $t_k$ (e.g. by an interferometric technique).  
We can explore this numerically by plotting $t_k$ as a function of $k\in(-\infty,+\infty)$ in the complex plane. Fig.\ \subref*{fig_1a_tk} shows the trajectory of $t_k$ for the case of $N=1$ with various parameters $\frac{\Gamma}{\Gamma^{{\rm tot}}}=1, 0.5,0.2$, where $\Gamma^{tot}=\Gamma+\Gamma'$. The trajectories go in the counter-clockwise direction when $k$ increases. The figure implies that $t_k$ encloses the origin when there is no bound state ($\frac{\Gamma}{\Gamma^{{\rm tot}}}>\frac{1}{2}$) and does not enclose the origin when the bound state exists ($\frac{\Gamma}{\Gamma^{{\rm tot}}}<\frac{1}{2}$). At the threshold ratio $\frac{\Gamma}{\Gamma^{{\rm tot}}}=0.5$, the trajectory passes through $0$ at $k=E_B=\omega_{eg}$.   The single-atom case shows that the winding number of $t_k$ around the origin is related to $N_B$.   Using Eq.\ \eqref{eq_tk} and the definition of the winding number, we find
\begin{gather}
\frac{1}{2\pi i}\int_{-\infty}^{+\infty}dk\ t_k^{-1}\frac{dt_k}{dk}=N-N_B, \label{eq:lev}
\end{gather}
where $N_B$ includes the number of bound states inside the continuum \cite{FN_Lev}.

To further illustrate the significance of Eq.\ \eqref{eq:lev}, we consider a two-atom system with interaction strength $V_1=V_2,$  distance $r_1-r_2=\frac{2\pi}{\omega_{eg}}$, and $\Gamma=\frac{V^2}{2}$. The dipole-dipole interaction induced by the 1D channel is 
$K= \Gamma \begin{pmatrix}
-i& -2i \\
  0 &-i
   \end{pmatrix}.
$
We consider the scenario when both atoms decay to the reservoir with rate $-i\Gamma'$ and interact coherently with strength $-\Gamma'$. In this case, 
$
K'=\Gamma' \begin{pmatrix}
-i& -1 \\
 -1&-i
   \end{pmatrix}
$\cite{FN_Kp}. 
 The total decay rate of each atom is $\Gamma^{{\rm tot}}=\Gamma+\Gamma'$.  Fig.\ \subref*{fig_2a_tk} shows the trajectory of $t_k$ for $\Gamma/\Gamma^{{\rm tot}}=0.2,\ 0.65,\ 0.75$.  All three trajectories are asymmetric about the real axis  \cite{FN_sym}. The corresponding number of bound states $N_B=2,\ 1,\  0$ decreases with decreasing dissipation. The corresponding winding number of $t_k$ increases from $0$ to $2$.

In many experiments, achieving $\Gamma/\Gamma^{{\rm tot}}$ above a few percent is difficult. Therefore, it is hard to achieve $N_B\neq N$ in the $N=1,2$ examples discussed above. However, we show in Ref.\ \cite{upcoming} that when $N$ is large,  $N_B<N$ can be achieved for realistic values of 1D decay rate $\Gamma$.

\emph{Diverging $g^{(2)}(0)$.---}Zeros in the transmission coefficient $t_{k}$ also lead to divergent bunching in the second-order photon-photon correlation function, which is an important signature of strongly-correlated light measured in the experiment. For infinitesimally weak coherent-state inputs, 
$g^{(2)}(\tau)=\frac{|\psi^{(2)}(r=\tau)|^2}{|t_k|^4}$, where $\psi^{(2)}(r)$ is the two-photon steady state  at the output and $r$ is the relative coordinate of the two photons \cite{SM_bs}.  The two-photon delay $\tau$ is equal to $r$ in units where the speed of light $c=1$.   When $t_k=0$, $\psi^{(2)}(r)$ is nonzero due to two-photon interactions mediated by the hard-core atomic interaction $U$.  Therefore, $g^{(2)}(\tau) =\infty$ at all $\tau$.  In the Supplemental Material \cite{SM_bs}, we calculate and plot $g^{(2)}(\tau)$ for $N=1$.

\emph{Derivation of spin model $M$.---}In this section, we derive $M$ and its relation to bound states.
Define the right bound state corresponding to energy $E_\alpha$:
\begin{gather}
|\psi_\alpha\rangle=\int_{-\infty}^{+\infty}dz \phi _\alpha(z)C^{\dagger}(z)|0,g\rangle +\sum_{j=1}^Ne_{j,\alpha}b_j^\dagger|0,g\rangle,  \label{eq_psialpha}
\end{gather}
where $\phi_\alpha(z)$ is the photon wavefunction and $e_{j,\alpha}$ is the excitation amplitude for the $j$-th  atom.
The left bound state and its normalization are discussed in the Supplemental Material \cite{SM_bs}. By definition, $H_1|\psi_\alpha\rangle=E_{\alpha}|\psi_\alpha\rangle$.  We integrate the steady-state equations of motion for photons along $z$ and get
\eq{
 \phi_{\alpha}(z)=\sum_{j=1}^Ne_{j,\alpha}V^*_j G_{E_\alpha}(z-z_j), \label{eqphi}
 }
where the coordinate-space free-photon propagator $ G_{\omega}(z')=-\eta_\omega i \Theta(\eta_\omega z')\exp(i\omega z')$ depends on $\eta_\omega$, the sign of Im[$\omega]$.  There is no inhomogeneous term in Eq.\ \eqref{eqphi} because the bound-state wavefunction vanishes at $z=\pm \infty$.

To solve for $e_{j,\alpha}$, Eq.\ \eqref{eqphi} is  substituted into the steady-state equations of motion for the atoms. We get 
\eq{
 E_\alpha e_{i,\alpha} \!=\! \sum_{j=1}^N (\omega_{eg}\delta_{ij} \!+\! K'_{ij} \!+\! V_iV^*_jG_{E_\alpha}(z_i\!-\!z_j))e_{j,\alpha}. \label{eqspine}
}
This equation can be reduced to an eigenvector calculation of an $N\times N$ matrix if $G_{E_\alpha}(z_i-z_j)$ is independent of $E_{\alpha}$. Let us study when this approximation can be made. 
When $\text{Im}[E_{\alpha}]<0$, $G_{E_{\alpha}}(z_i-z_j)=G_{\omega_{eg}-i0}(z_i-z_j)\exp(i(E_{\alpha}-\omega_{eg})(z_i-z_j))$, where $i0$ represents an infinitesimal imaginary number above the real line.  
Since $E_\alpha-\omega_{eg}$ is on the order of $\Gamma\sim \frac{V^2_i}{2}$, the phase $\exp(i(E_{\alpha}-\omega_{eg})(z_i-z_j))$ is negligible when the length of the 1D atomic cloud is much smaller than $c/\Gamma$. In this case, we can let $G_{E_\alpha}(z_i-z_j)= G_{\omega_{eg}-i0}(z_i-z_j)$, which corresponds to the Markov approximation.  Similarly, when Im$[E_\alpha]>0$, $G_{E_\alpha}(z_i-z_j)= G_{\omega_{eg}+i0}(z_i-z_j)$ under the Markov approximation.

Note that $K_{ij}$ defined in Eq.\ \eqref{eq_K1} is equal to  $V_iV^*_jG_{\omega_{eg}+i0}(z_i-z_j)$, and $K_{ij}^\dagger$ is equal to $V_iV^*_jG_{\omega_{eg}-i0}(z_i-z_j)$.  $G_{\omega_{eg}+i0}(z)$ and $G_{\omega_{eg}-i0}(z)$ are the retarded and advanced Green's functions of the free photons in the 1D channel. When Im$[E_\alpha]\leq 0$, Eq.\ \eqref{eqspine} becomes $E_{\alpha}e_{i,\alpha}=M_{ij}e_{j,\alpha}$, where $M$ is defined in Eq.\ \eqref{eq_M}. A self-consistency condition indicates that all the eigenvalues of $M$ below the real line in the complex plane are bound-state eigenvalues. It is easy to verify that the corresponding photon wavefunctions $\phi_{\alpha}$ calculated using Eq.\ \eqref{eqphi} vanish for large $z$. 

Does there exist a bound-state energy with positive imaginary component? If the answer is yes, $E_\alpha$ should be the eigenvalues of $M^{{\rm tot}}=\omega_{eg}+K'+K$ above the real line. However, as $M^{{\rm tot}}$ is dissipative, all its eigenvalues are either real or below the real line in the complex plane [See Fig.\ \subref*{fig:EigMtot}]. Hence there is no bound-state energy above the real line.

\textit{Outlook.---}Many of our findings can be generalized to multi-photon scattering processes, non-Markovian or multi-channel systems, other level structures and dispersion relations \cite{upcoming}.  Re-examining similar bound states in various dissipative systems may open the door to new insights into well-studied systems. Finally, with the high degree of control available in atomic, molecular, and optical systems, the first experimental study of Levinson's theorem is likely not far away.

\begin{acknowledgments}
We are grateful to M.~F.~Maghrebi, J.~S.~Douglas, M. T. Manzoni, A.~S.~S\o rensen and A.~Deshpande for discussions. YW and AVG acknowledge funding from ARL CDQI, AFOSR, ARO MURI, NSF PFC at JQI, NSF QIS, ARO, NSF Ideas Lab on Quantum Computing, and the DoE ASCR Quantum Testbed Pathfinder program.
DEC acknowledges funding from the MINECO ``Severo Ochoa" Programme for Centres of Excellence in R\&D (SEV-2015-0522), MINECO Plan Nacional Grant CANS, ERC Starting Grant FOQAL, Fundacio Privada Cellex, and CERCA Programme/Generalitat de Catalunya.
\end{acknowledgments}

\renewcommand{\thesection}{S\arabic{section}} 
\renewcommand{\theequation}{S\arabic{equation}}
\renewcommand{\thefigure}{S\arabic{figure}}
\setcounter{equation}{0}
\setcounter{figure}{0}
\setcounter{secnumdepth}{1}

\begin{widetext}
\section*{Supplemental Material}
In Sec.\ \ref{section_lbs}, we calculate the left bound states and discuss the normalization of the bound-state wavefunctions. In Sec.\ \ref{section_trans}, we derive the key result, Eq.\ (10) in the main text, on the relation between the transmission coefficient $t_k$ and the two spin matrices $M$ and $M^{\rm tot}$. In Sec.\ \ref{section_expl}, we illustrate the energy conservation relation associated with the scattering states and the eigenvectors of $M^{\rm tot}$ and $M$.  In Sec.\ \ref{section_g2}, we discuss the second-order photon-photon correlation function measured at the output when the input state is a weak coherent-state pulse.

\section{Left bound states and wavefunction normalization \label{section_lbs}}
For the non-Hermitian single-excitation Hamiltonian $H_1$, its left and right eigenvectors are different. In the main text, we only discussed the right bound states. For completeness, let us discuss the left bound state  $|\pi_\alpha\rangle$ corresponding to bound-state energy $E_\alpha$:
\begin{gather}
|\pi_\alpha\rangle=\int_{-\infty}^{+\infty}dz \bar{\phi} _\alpha(z)C^{\dagger}(z)|0,g\rangle +\sum_{j=1}^N\bar{e}_{j,\alpha}b_j^\dagger|0,g\rangle, \label{eq_psialpha}
\end{gather}
where $\bar{\phi}_{\alpha}$ is the photon wavefunction and $\bar{e}_{j,\alpha}$ is the excitation amplitude for atom $j$.
$|\pi_\alpha\rangle$ is also the right eigenvector of the Hermitian conjugate of  $H_1$ with eigenenergy $E^*_{\alpha}$:
\eq{
H_1^{\dagger}|\pi_\alpha\rangle=E^*_{\alpha}|\pi_\alpha\rangle.
}
The left bound states can be calculated in a similar method as the right bound states discussed in the main text. Integrating the steady-state equation of motion along the real axis, we get
\eq{
\bar{ \phi}_{\alpha}(z)=\sum_{j=1}^N\bar{e}_{j,\alpha}V^*_j G_{E^*_\alpha}(z-z_j).\label{eqphibar}
 }
From Eq.\ \eqref{eqphibar} and Eq.\ (13) in the main text, we can see that the bound-state photon wavefunction is localized around each atom with a width given by $1/\text{Im}[E_\alpha]$. 
 Substituting Eq.\ \eqref{eqphibar} into the steady-state equations of motion for the atomic excitation $\bar{e}_{j,\alpha}$, we get 
 \begin{gather}
M^\dagger_{ij}\bar{e}_{j,\alpha}=E^*_{\alpha}\bar{e}_{i,\alpha}.
\end{gather}

This set of eigenvalue equations determines the atomic amplitude $\bar{e}_{i,\alpha}$ up to a normalization constant.  We want to choose the normalization of $\bar{e}_{i,\alpha}$ and $e_{i,\alpha}$ such that the orthogonality $ \langle \pi_\alpha|\psi_\beta\rangle=\delta_{\alpha\beta}$ is ensured. 
Let us take a look at the overlap between the left and right bound states:
\begin{gather}
 \langle \pi_\alpha|\psi_\beta\rangle=\sum_{j=1}^N \bar{e}^*_{j,\alpha}e_{j,\beta} + \int_{-\infty}^{+\infty}dz \bar{\phi}^*_{\alpha}(z)\phi_{\beta}(z).
  \end{gather}
In this equation, the overlap between the photon wavefunctions is
\begin{gather}
\int_{-\infty}^{+\infty}dz \bar{\phi}^*_{\beta}(z)\phi_{\alpha}(z)\sim \sum_{i,j=1}^NV_jV^*_i(z_i-z_j)\exp(iE_\beta z_j-iE_\alpha z_i)\sim |V|^2L, 
\end{gather} where $L$ is the length of the 1D atomic cloud and $V$ is the scale of the interaction strength. Under the Markov approximation, $|V|^2L\ll 1$, the overlap between the photon wavefunctions is negligible. 
Therefore, if we choose $e_{i,\alpha}$ ($\bar{e}_{i,\alpha}$) to be the components of the normalized right (left) eigenvectors of $M$, the orthogonality relation $ \langle \pi_\alpha|\psi_\beta\rangle \approx \sum_{j=1}^N \bar{e}^*_{j,\alpha}e_{j,\beta}=\delta_{\alpha\beta}$ is ensured.

In the main text, we discussed the case of a 1D photon channel coupled to a single atom located at $z_1=0$. 
We showed that there exists a bound state with energy $E_B=\omega_{eg}-i\Gamma'+i\Gamma$ when $\Gamma'>\Gamma$. 
Here we give the expressions for the normalized right and left bound states $|\psi_B\rangle$ and $|\pi_B\rangle$ . The atomic excitation $e_{1,B}=\bar{e}_{1,B}=1$, and the photon wavefunctions are
\eq{
\phi_B(z)&=-iV^*\exp(i\omega_{eg}z+(\Gamma'-\Gamma)z)\Theta(-z),\\
\bar{\phi}_B(z)&=iV^*\exp(i\omega_{eg}z-(\Gamma'-\Gamma)z)\Theta(z).
}
The wavefunctions decay exponentially in space with width $1/(\Gamma'-\Gamma)$. When the atom is decoupled from the 1D channel, $V=0$, then $\phi_B(z)=\bar{\phi}_B(z)=0$. In this case, the bound state is simply the single atom in its excited state with eigenenergy $E_B=w_{eg}-i\Gamma'$.

\section{The transmission coefficient $t_k$ and its relation to $M,M^{\rm tot}$\label{section_trans}}
In this section we prove Eq.\ (10) in the main text and show that it is valid beyond the Markov approximation. The exact relation we want to prove is 
\eq{
t_k=\frac{\text{det}(k\mathbb{1}-M(k))}{\text{det}(k\mathbb{1}-M^{\rm tot}(k))}, \label{eq_t_k_fd}
}
where $M(k)=\omega_{eg}\mathbb{1}+K^\dagger (k)+K'$ and $M^{\rm tot}(k)=\omega_{eg}\mathbb{1}+K(k)+K'$. Here, $K_{ij}(k)=-iV_iV^*_j\exp(ik(z_i-z_j))$ is the frequency-dependent interaction between atoms $i$ and $j$ induced by the 1D photon channel without making the Markov approximation. 

To calculate $t_k$, we solve for the right scattering-state wavefunction corresponding to energy $k$.
Integrating the steady-state equation of motion at energy $k$ and choosing the boundary condition $\lim_{z\rightarrow -\infty}\phi_k(z)=\exp(ikz)$, we get 
\eq{
 \phi_{k}(z)=-i\sum_{j=1}^Ne_{j,k}V^*_j \exp(ik(z-z_j))+\exp(ikz), \label{Seqphi}
 }
 where $e_{j,k}$ is the atomic excitation defined in Eq.\ (9) in the main text. This gives the transmission coefficient 
 \eq{
t_k=1-i\sum_{j=1}^Ne_{j,k}V^*_j \exp(-ikz_j).\label{eqtk}
}
Substituting Eq.\ \eqref{Seqphi} into the steady-state equation of motion for the atoms, we obtain a set of linear equations for $e_{j,k}$:
\eq{
(k\delta_{ij}-M^{\rm tot}_{ij}(k))e_{j,k}=V_i\exp(ikz_i).\label{eqek}
}

To prove Eq.\ \eqref{eq_t_k_fd}, we collect the atomic-excitation amplitudes and the interaction amplitudes by defining two $N$-dimensional vectors:
\eq{
|e_k\rangle &\equiv (e_{1,k},\dots , e_{N,k})^T,  \label{eq_ek}\\
|v_k\rangle& \equiv (V_1\exp(ikz_1),\dots ,V_N\exp(ikz_N))^T. \label{eq_vk}
}
Now, Eq.\ \eqref{eqek} can be rewritten as
\eq{
|e_k\rangle&=(k\mathbb{1}-M^{\rm tot}(k))^{-1}|v_k\rangle. \label{eq_ekvk}
}
And Eq.\ \eqref{eqtk} can be rewritten as
\eq{
t_k&=1-i\langle v_k|(k\mathbb{1}-M^{\rm tot}(k))^{-1}|v_k\rangle\\
&=1-i\text{Tr} [|v_k\rangle\langle v_k| (k\mathbb{1}-M^{\rm tot}(k))^{-1}]. \label{eq:tkv}
}

Using  the property $K(k)-K^\dagger(k)=-i|v_k\rangle \langle v_k|$, we get $M(k)=M^{\rm tot}(k)+i|v_k\rangle \langle v_k|$, and 
Eq.\ \eqref{eq_t_k_fd} can be rewritten as
\eq{
t_k=\frac{\text{det}(k\mathbb{1}-M^{\rm tot}(k)-i|v_k\rangle\langle v_k|)}{\text{det}(k\mathbb{1}-M^{\rm tot}(k))}.\label{eq:tk_deco}
}
Our goal is to prove that Eq.\ \eqref{eq:tkv} is equivalent to Eq.\ \eqref{eq:tk_deco}. 
For simplicity of notation, let us define $X=k-M^{\rm tot}(k)$ and $|v\rangle=|v_k\rangle$. The equality we want to prove is reduced to 
\eq{
\frac{\text{det}(X-i|v\rangle\langle v|)}{\text{det}(X)}=1-i\text{Tr}(|v\rangle\langle v|X^{-1}).\label{eqTr_det}
}
Since determinant and trace are basis-independent, we can choose an orthonormal basis such that $|v\rangle$ is parallel to one of its basis vectors. Without loss of generality, we can choose $|v\rangle=|v|\cdot (1,0,\dots, 0)^T$, where $|v|$ is a normalization factor. With this choice of basis, there is only one non-zero matrix element in the matrix $-i|v\rangle\langle v|$:
\eq{
-i|v\rangle\langle v|=-i|v|^2\left(\begin{matrix}
1 & 0 &\dots &0\\
0 & 0 &\dots &0 \\
\vdots\\
 \end{matrix}\right). \label{eq_vv}
 }
The simple structure of Eq.\ \eqref{eq_vv} simplifies the calculation of both sides of Eq.\ \eqref{eqTr_det}. The right-hand side of Eq.\ \eqref{eqTr_det} is
\eq{
1-i\text{Tr} [|v\rangle\langle v| X^{-1}]=1-i|v|^2(X^{-1})_{11},\label{eq_lhs}
}
where $(X^{-1})_{11}$ is the $(1,1)$ matrix element of $X^{-1}$. The left-hand side of Eq.\ \eqref{eqTr_det} is 
\eq{
\frac{\text{Det}(X-i|v\rangle\langle v|)}{\text{Det}(X)}=\frac{\text{Det}(X)-i|v|^2 L_{11}}{\text{Det}(X)}, \label{eq_rhs}
}
where $L_{11}=\text{Det}(X_{\cancel{1},\cancel{1}})$, and $X_{\cancel{1},\cancel{1}}$ is the minor matrix of $X$ obtained by eliminating the $1^ {st}$ row and  $1^{st}$ column of $X$. Using the property of the inverse 
\eq{
(X^{-1})_{11}=\frac{L_{11}}{\text{Det}(X)},
}
we can see that the right-hand side of Eqs.\ \eqref{eq_rhs} and \eqref{eq_lhs} are equal, which concludes the proof.\\
\section{Energy exchange processes associated with $M^{\rm tot}$ and $M$\label{section_expl}}
\begin{figure}{}
\subfloat[$M^{\rm tot}|\tilde{e}_{\alpha}\rangle=\tilde{e}_{\alpha}|\tilde{e}_{\alpha}\rangle$]{
    \includegraphics[width=0.16\linewidth]{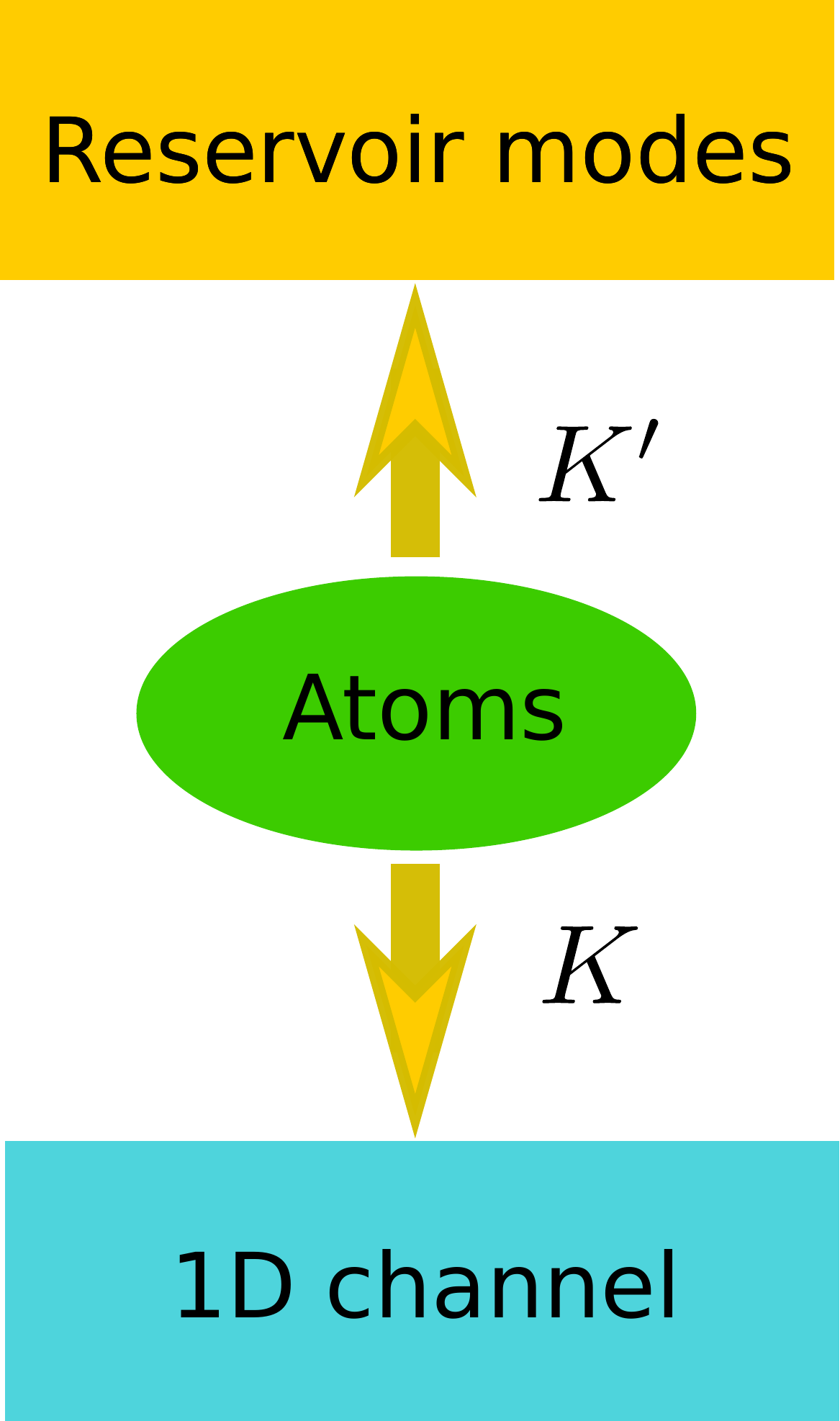} \label{fig_procMtot}
}
\hspace{20mm} 
 \subfloat[ $M|e_{\alpha}\rangle=e_{\alpha}|e_{\alpha}\rangle$]{
    \includegraphics[width=0.16\linewidth]{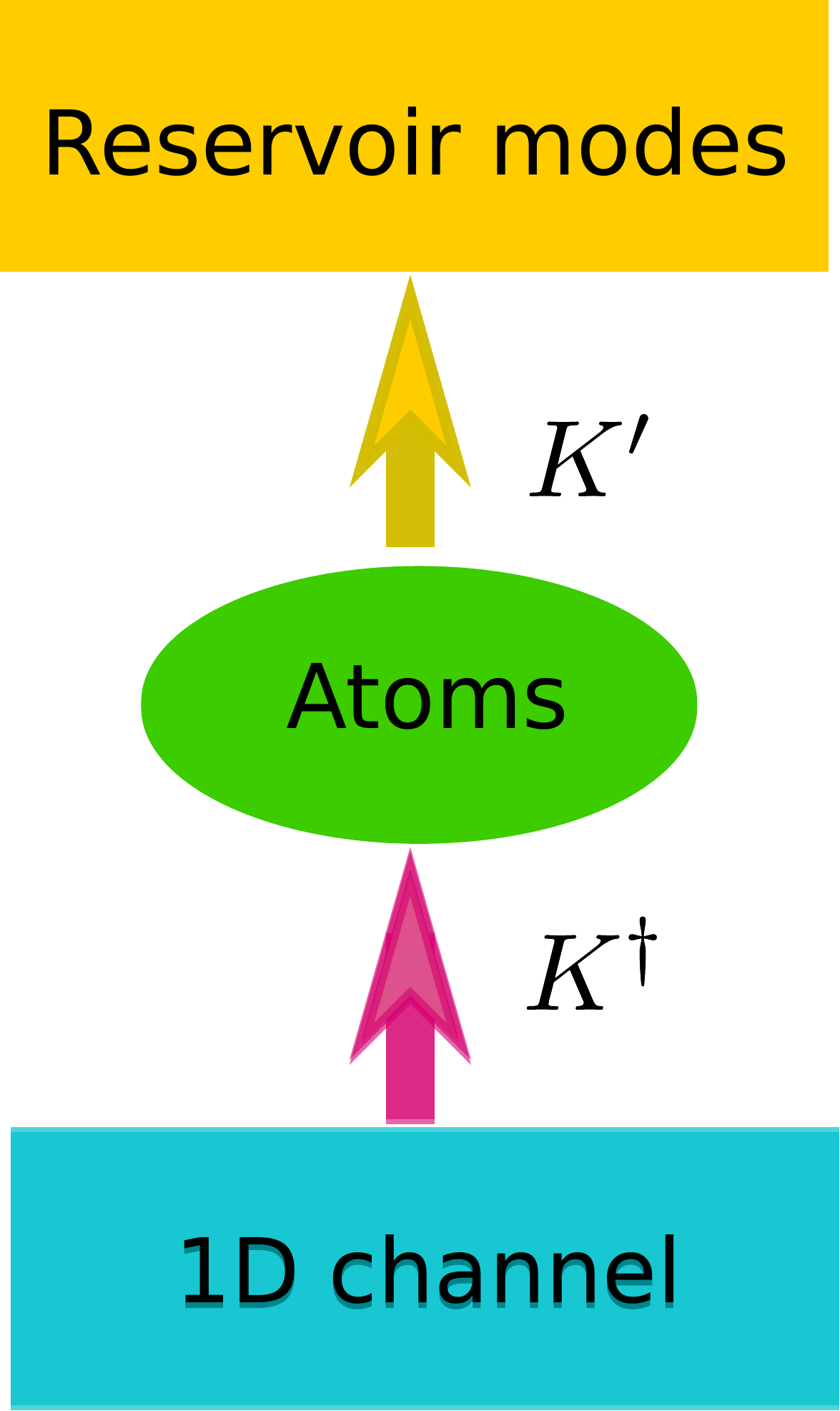} \label{fig_procM}
  }
 \hspace{10mm} 
  \subfloat[ $(k\mathbb{1}-M^{\rm tot})|e_k\rangle=|v_k\rangle$]{
    \includegraphics[width=0.243\linewidth]{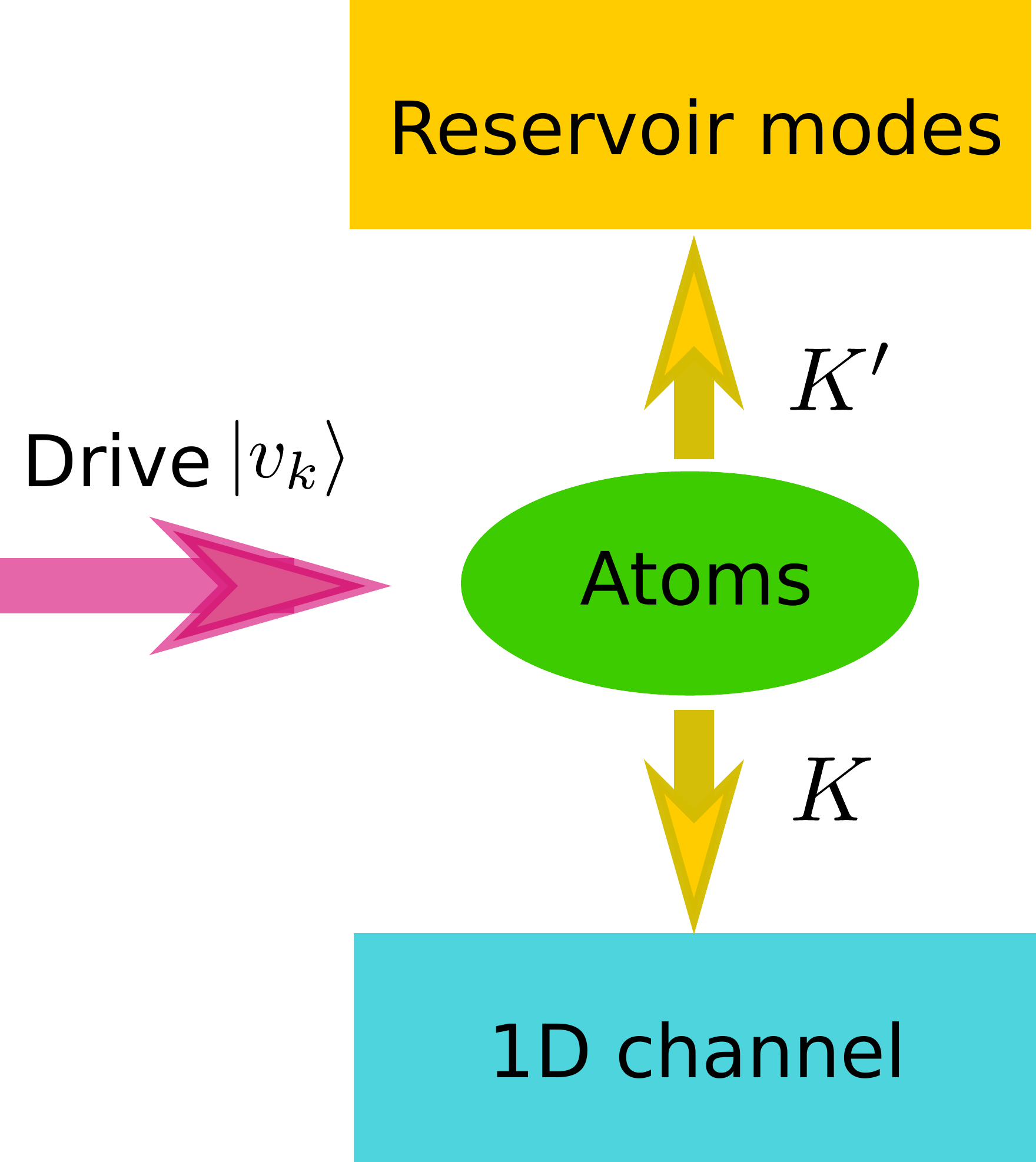} \label{fig_proc_scc}
}

\caption{The energy flow digrams associated with (a) the eigenvectors of $M^{\rm tot}$, (b) the single-photon bound states  and (c) the scattering states.  $M^{\rm tot}=w_{eg}\mathbb{1}+ K+K'$ and $M=w_{eg}\mathbb{1}+ K^\dagger+K'$. }\label{fig_proc}
\end{figure}

In this section, we illustrate in Fig.\ \ref{fig_proc} the energy exchange processes associated with the eigenvectors of $M^{\rm tot}$, $M$, and the scattering states.  To be specific, we want to compare the following three equations:
\eq{
M^{\rm tot}|\tilde{e}_{\alpha}\rangle&=\tilde{e}_{\alpha}|\tilde{e}_{\alpha}\rangle, \label{eq_Mtoteig}\\
M|e_{\alpha}\rangle&=e_{\alpha}|e_{\alpha}\rangle,  \label{eq_Meig}\\
(k\mathbb{1}-M^{\rm tot})|e_k\rangle&=|v_k\rangle.  \label{eq_scattering}
}
Eqs.\ \eqref{eq_Mtoteig} and \eqref{eq_Meig} are the eigenequations of $M^{\rm tot}$ and $M$ with respective eigenvectors $|\tilde{e}_{\alpha}\rangle=(\tilde{e}_{1,\alpha},\dots, \tilde{e}_{N,\alpha})^T$ and $|e_{\alpha}\rangle=(e_{1,\alpha},\dots, e_{N,\alpha})^T$.  Eq.\ \eqref{eq_scattering} is equivalent to Eq.\ \eqref{eq_ekvk},  which is the linear equation for the calculation of atomic amplitudes for the scattering states. The vectors that represent the atomic amplitudes $|e_k\rangle$ and the drive $|v_k\rangle$ were defined in Eqs.\ \eqref{eq_ek} and \eqref{eq_vk}.

Since $M^{\rm tot}$ is the single-excitation effective Hamiltonian matrix when all photons are traced out of the system, its eigenvector $|\tilde{e}_\alpha\rangle$ corresponds to a state $|\tilde{\psi}_{\alpha}\rangle=\sum_{i=1 }^N \tilde{e}_{i,\alpha} b^\dagger_i|0,g\rangle $ that decays to all photon channels during time evolution without changing shape.  
Fig.\ \subref*{fig_procMtot} illustrates the direction of the energy flow from the atoms during the time evolution of $|\tilde{\psi}_{\alpha}\rangle$,  where the atoms dissipate energy to the 1D and reservoir modes through dissipative interactions $K$ and $K'$, respectively. The total energy loss rate of the atoms is given by Im$[\tilde{e}_\alpha]$.

The eigenvectors $|e_\alpha\rangle$ of $M$ correspond to single-photon bound states $|\psi_\alpha\rangle $ when Im$(e_\alpha)<0$.  These photon-atom hybrid states dissipate energy to the reservoir modes without changing shape.
Fig.\ \subref*{fig_procM} illustrates the direction of energy flow between the atoms and photon channels during the time evolution of $|\psi_{\alpha}\rangle$. The atoms absorb energy from the 1D channel through interaction $K^\dagger$ and dissipate energy to the reservoir modes through $K'$. The net rate of energy change for the atoms is Im$[e_\alpha]<0$. 

For a scattering state $|\psi_k\rangle$ with atomic amplitudes given by the vector elements of $|e_k\rangle$, the direction of energy exchange is illustrated in Fig.\ \subref*{fig_proc_scc}. The atoms absorb energy through the drive (represented by vector $|v_k\rangle$) and dissipate energy to all photon channels through dissipative interaction $K+K'$. The magnitude of the atomic amplitudes is constant during time evolution, as the energy-absorption rate from the drive is equal to the energy-dissipation rate to the photon channels. 

\section{Second-order photon-photon correlation function for a weak coherent pulse \label{section_g2}}

In this section, we give the definition of the second-order photon-photon correlation function $g^{(2)}(\tau)$ and calculate $g^{(2)}(\tau)$ for the output of a continuous weak coherent input scattered by a single atom. To solve this problem, we consider a long weak coherent pulse of uniform amplitude. 

We first define the creation operator $ \mathcal{E}^\dagger$ which generates a single photon with duration $T$ and center $z_0$ at the initial time $t=0$:
\eq{
 \mathcal{E}^\dagger&=\frac{1}{\sqrt{cT}}\int_{-\infty}^{+\infty} dz l(z) \exp(ikz)C^\dagger(z),\\
 l(z)&=\Theta(z-z_0+cT/2) \Theta (-z+z_0+cT/2).
 }
 $\mathcal{E}^{\dagger}$ and its Hermitian conjugate $\mathcal{E}$ satisfy the commutation relation $[\mathcal{E},\mathcal{E}^\dagger]=1$. 
The coherent-state pulse with average photon number $|\alpha|^2$ is defined as
\eq{
 |\alpha\rangle=\exp(-|\alpha|^2/2)\left( \sum_{n=0}^\infty \frac{\alpha^n}{\sqrt{n!}}\mathcal{E}^{\dagger n}|0\rangle\right),\label{eq_alphaT}
 }
 where $|0\rangle$ is the vacuum state and $\exp(-|\alpha|^2/2)$ is the normalization factor.  
$|\alpha\rangle$ is prepared at time $t=0$, sent through the atomic cloud, and measured at time $t_f$ after the pulse has completely exited the cloud. Because of the dissipation induced by the reservoir photon modes, the output pulse at $t_f$ is described by a density matrix $\rho$.

The second-order photon-photon correlation function for the output pulse can be calculated using
  \eq{
 g^{(2)}(r_1,r_2)&=\frac{\langle C^\dagger(r_1) C^\dagger(r_2) C(r_1) C(r_2)\rangle}{\langle C^\dagger (r_1)C (r_1)\rangle \langle C^\dagger (r_2)C (r_2)\rangle},\label{g2n}
 }
   where the coordinates $r_1$ and $r_2$ of the measurements are  chosen within the length of the output pulse away from the edges, and the averages are taken with respect to the density matrix $\rho$. 
   
 $\Gamma \sim V_i^2/2$ is the energy scale of the atom-photon interaction and $\mathcal{R}=|\alpha|^2/T$ is the rate of the incoming photons. When  $\mathcal{R}/\Gamma\ll 1$,  the probability of having two photons simultaneously interacting with the atoms is much smaller than $1$. In the Supplemental Material of Ref.\ \cite{Liang2018}, it is shown that, when  $\mathcal{R}/\Gamma\ll 1$,  the density of photons $\langle C^\dagger (r)C (r)\rangle$ and the two-point correlation function $\langle C^\dagger(r_1) C^\dagger(r_2) C(r_1) C(r_2)\rangle $ evaluated with respect to the output state $\rho$ depend only on the single-photon and two-photon scattering processes, respectively. Specifically, in the limit $\mathcal{R}/\Gamma \rightarrow 0$,

 \eq{
 \langle C^\dagger(r)C(r)\rangle &= \mathcal{R}|t_k|^2 ,\\
  \langle C^\dagger(r_1)C^\dagger(r_2)C(r_1)C(r_2)\rangle &=\mathcal{R}^2|\psi^{(2)}(r_1,r_2)|,
 }

where $t_k$ is the single-photon transmission coefficient. $\psi^{(2)}(r_1,r_2)$ is the coordinate-space wavefunction of  $|\psi^{(2)}\rangle$, which is the two-photon output state corresponding to the input state $|k,k\rangle$ of two free photons with frequency $k$. $|\psi^{(2)}\rangle=S|k,k\rangle$ where $S$ is the two-photon S-matrix.  Here, we have assumed that the bandwidth of the pulse satisfies $1/T\ll \Gamma$, so  the scattering amplitudes of the different number-state manifolds of the pulse are very close to those of the plane waves. 

We define the center of mass coordinate $R=(r_1+r_2)/2$ and the relative coordinate $r=r_1-r_2$.  Due to the exchange symmetry of photons, $\psi^{(2)}(r,R)=\psi^{(2)}(-r,R)$. Except for the edges of the two-photon wavefunction, the output satisfies $\psi^{(2)}(r,R)=\exp(2ikR)\psi^{(2)}(r)$. Therefore, $g^{(2)}(r,R)$ is independent of the center of mass coordinate $R$ and $g^{(2)}(r)=g^{(2)}(-r)$. For speed of light $c=1$, the two-photon delay $\tau$ is equal to $r$. In the limit $\mathcal{R}/\Gamma\rightarrow 0$, Eq.\ \eqref{g2n} becomes
  \eq{
 g^{(2)}(\tau)&=\frac{|\psi^{(2)}(r=\tau)|^2}{|t_k|^4}.\label{g2rn}
 }
When $t_k\rightarrow 0$ and $\psi^{(2)}(r)$ is finite, $g^{(2)}(\tau=r)\rightarrow \infty$. 
Next, we give the details of the calculation of $g^{(2)}(\tau)$ at the output when a pulse of weak coherent state  is scattered by a single atom. To calculate $\psi^{(2)}(r)$ in Eq.\ \eqref{g2rn}, we refer to Refs.\ \cite{Shi2015, Rephaeli2013} for the expression of the symmetrized two-photon S-matrix when the energies of the two incoming photons are $k_1, k_2$ and the energies of the output photons are $k'_1, k'_2$. In the center of mass frame, $E=k_1+k_2$, $q=(k_1-k_2)/2$, $E'=k'_1+k'_2$, $q'=(k'_1-k'_2)/2$. $S(E',q',E,q)\equiv \langle k'_1,k'_2|S|k_1, k_2\rangle$ is given by
\begin{gather}
S(E',q',E,q)=t_{E/2+q}t_{E/2-q}(\delta(q-q')+\delta(q+q'))\delta(E-E')-4\pi i T(E',q',E,q)\delta(E-E'),
\end{gather}
where the T-matrix element $T(E',q',E,q)$ is given by 
\begin{gather}
T(E',q',E,q)
=-\frac{16\Gamma^2}{\pi^2}\frac{E-2w_{eg}+2i\Gamma^{\rm tot}}{[4q^2-(E-2w_{eg}+2i\Gamma^{\rm tot})^2][4{q'}^2-(E-2w_{eg}+2i\Gamma^{\rm tot})^2]}.
\end{gather}
$\Gamma=V^2/2$ is the decay rate of the atom to the 1D channel and $\Gamma^{\rm tot}=\Gamma'+\Gamma$ is the total decay rate of the atom. 
The first term of $S(E',q',E,q)$ is equal to the amplitude of two photons scattering off the atom independently. The second term is the amplitude of scattering processes involving the non-linearity of the atom. $\psi^{(2)}(r,R)$ can be computed from a Fourier transform of $S(E',q',E, q=0)$ with respect to $E',q'$. For a weak coherent-state pulse resonant with the two-level atom, the two input photons have the same energy $k_1=k_2=w_{eg}$, so $\psi^{(2)}(r,R)$ is given by
\eq{
\psi^{(2)}(r,R)&=\frac{1}{\sqrt{2\pi}}\int_{-\infty}^{+\infty} dq'dE' \exp(iq'r+iE'R) S(E',q',E=2w_{eg}, q=0)\\
&=\left[\frac{(\Gamma'-\Gamma)^2}{\pi(\Gamma^{\rm tot})^2}\cos(qr)-\frac{4\Gamma^2}{\pi(\Gamma^{tot})^2} \left(\exp(\Gamma^{\rm tot}r)\Theta(-r)+\exp(-\Gamma^{\rm tot}r)\Theta(r)\right)\right]\exp(2iw_{eg}R).\label{eq_psi2}
}
The transmission coefficient on resonance is $t_{k=w_{eg}}=\frac{\Gamma'-\Gamma}{\Gamma^{\rm tot}}$. Using Eq.\ \eqref{g2rn} and  Eq.\ \eqref{eq_psi2}, we plot $g^{(2)}(\tau)$ for $N=1$ as a function of  $\Gamma/\Gamma^{\rm tot}$. When $\frac{\Gamma}{\Gamma^{\rm tot}}=0.5$, the single-photon transmission $t_{k=w_{eg}}=0$ and $g^{(2)}(\tau)$ diverges. 
\begin{figure}[t]
\includegraphics[width=0.4\linewidth]{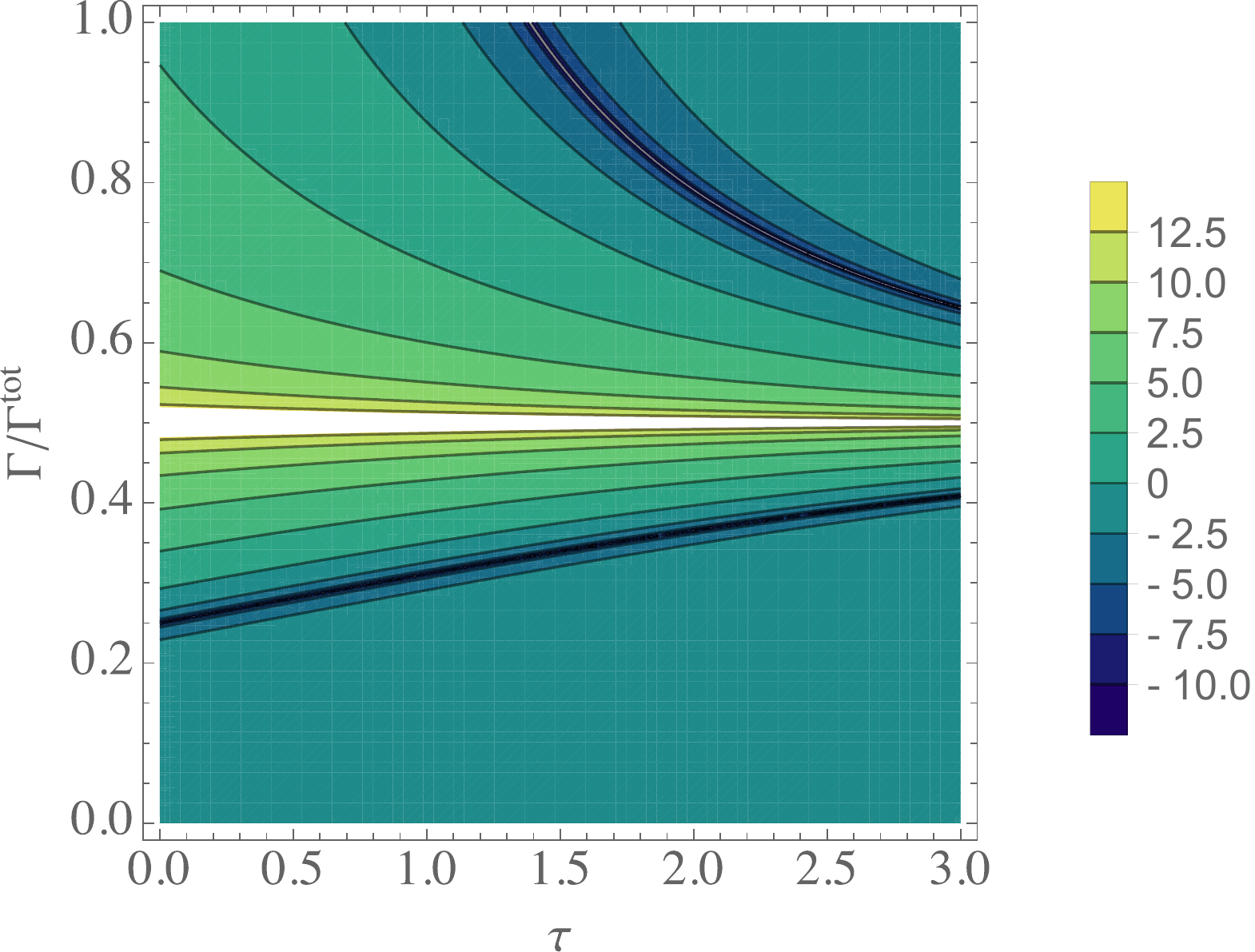}
\caption{(Color online) log$(g^{(2)}(\tau))$ as a function of the ratio $\Gamma/\Gamma^{\rm tot}$ and $\tau=r$. At $\Gamma/\Gamma^{\rm tot}=0.5$, $t_{k=w_{eg}}=0$ and $g^{(2)}(\tau)$ diverges for all $\tau$.}
\label{fig_g2_3d}
\end{figure}

\end{widetext}


\begin{thebibliography}{59}
\bibitem [{FN_({\natexlab{b}})}]{FN_Kp}%
  \BibitemOpen
\bibinfo {note} {{The matrix $K$ for any
  $N$-atom system satisfies the property that $K-K^\dagger=-i|v\rangle \langle
  v|$ is proportional to a projector in the $N$-dimensional space \cite{SM_bs}.
  In principle, $K'$ can be any $N\times N$ dissipative matrix }}\BibitemShut
  {NoStop}%
\bibitem [{FN_({\natexlab{c}})}]{FN_sym}%
  \BibitemOpen
 \bibinfo {note} {{ The trajectory of
  $t_k$ is symmetric about the real axis when the eigenvalues of $M$ and
  $M^{\rm tot}$ are symmetric about the same vertical line Re$[E]=E^*$ in the
  complex plane (Re$[E]$,Im$[E]$) for some real number $E^*$ }}\BibitemShut
  {NoStop}%
\end{thebibliography}
\end{document}